\documentclass[aps,prl,twocolumn,superscriptaddress]{revtex4}
\usepackage{graphicx}%        Include       figure       files
\usepackage{amsmath,amssymb}

\newcommand{\fii}{\varphi}

\begin{document}

\title{Solitonization of the Anderson Localization}

\author{Claudio Conti} 
\affiliation{Department of Physics, Univiversity Sapienza, Piazzale Aldo Moro 5, 00185 Rome, Italy}
\date{\today}
%\pacs{...}

\begin{abstract}
We study the affinities between the shape of the bright soliton of the one-dimensional nonlinear Schroedinger equation
and that of the disorder induced localization in the presence of a Gaussian random potential.
With emphasis on the focusing nonlinearity, we consider the bound states of the nonlinear Schroedinger equation
with a random potential; for the state exhibiting the highest degree of localization, we derive explicit expressions for the
nonlinear eigenvalue and for the localization length by using perturbation theory and a variational approach 
following the methods of statistical mechanics of disordered systems. 
We numerically investigate the linear stability and ``superlocalizations''. The profile of the disorder averaged Anderson localization
is found to obey a nonlocal nonlinear Schroedinger equation.
\end{abstract}
\maketitle

%%%%%%%%%%%%%%%%%
\noindent {\it Introduction ---}
Solitons \cite{ZK65}, and disorder induced Anderson states \cite{Anderson58} are two apparently unrelated forms of wave localization, 
the former being due to nonlinearity \cite{DrazinBook}, the latter to linear disorder \cite{LifshitsBook}.
However, on closer inspection, they look similar for various reasons:
they are exponentially localized, they correspond to appropriately defined negative eigenvalues, 
they may be located in any position in space (which is homogeneous for solitary waves, and populated
 by a random potential for Anderson states).
Furthermore, various recent investigations deal with the theoretical, numerical and experimental analysis of localized states in 
the presence of disorder and nonlinearity \cite{Kivshar1990, Zavt93, Sukh2001,Staliunas03, bliokh06, bliokh08,  Kivshar10, Folli10, Denz2011, Folli11,Folli12, Skupin2012, Sacha09, kartashov08, Modugno2010},
as specifically in optics \cite{Swartz07,Conti08PhC, Lagendijk2012}, in Bose-Eistein condensation (BEC) \cite{Shapiro2010,Paul2007,AspectNature2008,InguscioNature2008}, and more recently in for random lasers \cite{Leonetti2012}.
The nonlinear Anderson localizations are expected to have a power (number of atoms for BEC, pump fluence for random lasers or active cavities) dependent localization length, 
and eigenvalue, but also exist for a vanishing nonlinearity: in the low fluence regime, they are Anderson localizations, but
at high fluence, it is expected that they are more related to solitons.
Such a situation resembles other forms of linear localization, as the multidimensional ``localized waves'' \cite{RecamiBook},
which are ``dressed'' by the nonlinearity \cite{Conti04c};
a key difference with respect to localized waves is that Anderson states are square-integrable, another feature in common with bright solitons \cite{KivsharBook}.
\\\noindent Many authors investigated the effect of nonlinearity on Anderson localization, as, e.g., \cite{Gred92, Kopidakis2000,Pikovsky2008,Flach2009,Fishman2012},  
here we report on a theoretical and numerical analysis that allows to derive explicit formulae describing the
nonlinear dressing of the fundamental Anderson states, and the way they become solitons as the nonlinear effects are dominant. 
We show that the disorder averaged profile of the nonlinear Anderson localization is given by the very same equation providing the soliton 
shape, augmented by a power and disorder dependent term. This equation defines a particular highly nonlocal nonlinear response \cite{Snyder97},
and results in quantitative agreement with computations. In addition, we numerically demonstrate that these states are stable with respect to small perturbations, and that this stability is driven by a novel kind of localization, which we address as ``superlocalization'', resulting from the interplay
of solitons and Anderson states.
\\\noindent {\it Outline ---}
We review the nonlinear Schroedinger equation with a Gaussian random potential; we describe the weak perturbation theory for small nonlinearity, when linear Anderson states are slightly perturbed; we consider the strong perturbation theory, i.e., the regime where the disordered potential is negligible and the only form of localization is the bright soliton;
we compare the two limits with numerical simulations; we use a phase-space variational approach to derive results valid at any order of nonlinearity and quantitatively in agreement with the two mentioned limits and with numerical analysis; the stability is finally numerically demonstrated.
\\\noindent {\it The model ---}
We consider the one-dimensional Schroedinger equation with random potential $V(x)$:
\begin{equation}
i\psi_t=-\psi_{xx}+V(x) \psi - \chi |\psi|^2 \psi\equiv \mathcal{N}[\psi]\text{,}
\label{nls1}
\end{equation}
where $\chi=1$ ($\chi=-1$) corresponds to the focusing (defocusing) case. 
$V(x)$ is Gaussianly distributed such that $\langle V(x)V(x') \rangle=V_0^2 \delta(x-x')$.  

\noindent The linear states ($\chi=0$) $\psi=\varphi\exp(-i E t)$ are given by \begin{equation}
-\varphi_{xx}+V(x) \varphi=\mathcal{L}\varphi=E \varphi\text{,}
\end{equation}
and we consider the lowest energy localized states with negative $E$; these are $\varphi_n(x)$ with $(\varphi_n,\varphi_m)=\delta_{nm}$ and $\delta_{nm}$ the Kronecker delta. 
We denote as $E_0$ the lowest negative energy
of the linear ($\chi=0$) fundamental state $\varphi_0$.
We stress that $\varphi_0$ is the fundamental linear state with unitary norm, and in the following we will use states such that $P=\int |\psi|^2 dx$ as the solutions of the nonlinear equation (\ref{nls1}); $P$ measures the strength of the nonlinearity 
as $\chi=\pm 1$ in the adopted scaling.

\noindent The localization length is calculated by the inverse participation ratio \begin{equation}
l=\frac{|\int \varphi^2 dx|^2}{\int \varphi^4 dx}=\frac{P^2}{\int \varphi^4 dx}\text{.}
\label{eq:locldef}
\end{equation}
\noindent For example, for an exponentially localized state $\varphi^{(e)}=\sqrt{P}\varphi_e$ with $\varphi_e=\exp{(-2|x|/\bar l)}/\sqrt{\bar l/2}$, we have $l=\bar l$.
$l_0$ is the linear localization length of the fundamental state $\sqrt{P}\varphi_0$.
We recall that for the linear problem, the number of states per unit length is known and the mean value for the
energy can be approximated by $\overline E_L\cong -V_0^{4/3}/3$ \cite{Halperin65,LifshitsBook}.
Here $\overline E_L$  is the mean linear negative value of the energy levels of a Gaussian random potential, and will be used below as the appropriate limit for $P\rightarrow 0$. We also recall that in the linear case, the energy scales like the inverse squared localization length \cite{LifshitsBook}, as also found below when $P\rightarrow 0$.

\noindent {\it The Lyapunov functional --- }
The nonlinear Anderson states for a specific disorder realization $V(x)$ are the solutions of
\begin{equation}
\mathcal{N}(\varphi)=-\varphi_{xx}+V(x) \varphi-\chi \varphi^3=E\varphi\text{,}
\label{eq:nlbound}
\end{equation}
which are obtained numerically, as detailed below.
The nonlinear states correspond to the extrema of the Lyapunov functional 
\begin{equation}
F=\int \{ |\psi_x|^2+[V(x)-E] |\psi|^2-\frac{\chi}{2}|\psi|^4 \} dx
\label{liap1}
\end{equation} with the Hamiltonian $H=F+E P$.
\\As $E=\left(\fii,\mathcal{N}[\fii]\right)/(\fii,\fii)=\left(\fii,\mathcal{N}[\fii]\right)/P$, one has, for the solutions of (\ref{eq:nlbound}), $F=\chi(\varphi^2,\varphi^2)/2$, that is
\begin{equation}
F=\frac{\chi}{2}\int \varphi^4 dx=\frac{\chi}{2}\frac{P^2}{l}\text{,}\hspace{1cm} \text{or}\hspace{1cm} \frac{1}{l}=\frac{2F}{\chi P^2}\text{,}
\label{eq:Fl}
\end{equation}
which show that a connection between the Lyapunov functional $F$ and the localization length exists.

\noindent{\it Weak perturbation theory ---}
For small $P$, standard perturbation theory \cite{Folli12} on $\sqrt{P}\varphi_0(x)$ gives 
\begin{equation}
E=E_0- \chi \frac{P}{l_0}+O(P^2)\text{,}
\label{eq:weakE}
\end{equation}
where $l_0$ is the {\it linear} localization length in Eq.(\ref{eq:locldef}). 
For $\chi=1$, $E<0$ decreases as the power is increased, while increases and eventually changes sign in the defocusing case ($\chi=-1$). 
Being $\varphi^{(1)}$ the standard first order correction to the linear state $\varphi_0$, we find at order $O(P)$ for the localization length:
\begin{equation}
\begin{array}{l}
l=\frac{\left(\varphi,\varphi\right)^2}{\left(\varphi^2,\varphi^2\right)}
\cong \frac{1}{\left(\varphi_0^2,\varphi_0^2\right)+
4P(\varphi_0^3,\varphi^{(1)})}= l_0 \left[1- 4 P \frac{\left(\varphi_0^3,\varphi^{(1)}\right)}{\left(\varphi_0^2,\varphi_0^2\right)}\right]=\\ =l_0-4 \chi P l_0^2 \sum_{n>0}\frac{\left(\varphi_n,\varphi_0^3\right)^2}{E_n-E_0}=
l_0\left(1-\chi\frac{P}{P_0}\right)
\label{lweak}
\end{array}
\end{equation}
Eq.(\ref{lweak}) predicts that $l$ increases (decreases) with $P$ in the defocusing (focusing) case; $P_0=\left[4 l_0 \sum_{n>0}(\varphi_n,\varphi_0^3)^2/(E_n-E_0))\right]^{-1}$ gives the power level such that, when $\chi=1$, $l$ vanishes, and this is defined as the critical power for the transition to a solitonic regime, where the weak expansion is expected not to be valid.
$P_0$ depends on the linear eigenstates of the potential and comes from the lowest order perturbation expansion of the localization length. 

Summarizing, the weak expansion allows to affirm that two critical powers can be defined: in the defocusing case, there is a power $P=|E_0| l_0$ at which the eigenvalue changes sign, corresponding to a nonlinearity that destroys the Anderson states; in the focusing case there is a power $P=P_0$ at which the localization length vanishes, this is the nonlinearity level needed to the bound state for resembling a bright soliton (i.e., for the ``solitonization'' of the Anderson state). In the weak expansion these critical powers are dependent on the specific disorder realization and have a statistical distribution. 
In a later section, we report a variational approach that allows to derive closed expressions for the peak of these distributions $P_C$, which depends only on $V_0$;
we will limit to the focusing case, as the defocusing one requires a separated treatment.

\noindent {\it Strong perturbation theory --- }
For large $P$, we write the solution by a {\it multiple scale expansion} as $\varphi=P \eta(P x)$ 
and Eq.(\ref{eq:nlbound}), at the highest order in $P$ ($x_P\equiv P x$, $E=\left(\varphi,\mathcal{N}[\varphi]\right)/P\equiv P^2 E_P$), reduces to
\begin{equation}
-\frac{d^2\eta}{d x_P^2}-\chi \eta^3=E_P \eta\text{,}
\label{eq:strong}
\end{equation}
where $E_P$ is the eigenvalue scaled by $P^2$.
For large $P$, the nonlinear Anderson states are asymptotically described
by the solitary-wave solutions in a manner substantially independent of $V(x)$. 
For $\chi=1$, Eq.(\ref{eq:strong}) is satisfied by the fundamental bright soliton \cite{DrazinBook} and, correspondingly, we have $\varphi=\sqrt{-2 E}/\cosh(\sqrt{-E} x)$. \\We stress that, in this expansion, $E<0$ as for the linear Anderson states,
being $E=E_S=-P^2/16$ and $l=l_S=12/P$, the subscript referring to the {\it soliton} trends.
$E_S$ is the ``nonlinear eigenvalue" for the soliton, which is determined by $P$, while $l_S$ is the corresponding localization length in this strong perturbation expansion, i.e., when neglecting the linear potential $V(x)$. Note also that this trend at high power is also expected for the higher order states $\varphi_n$.
%%%%%%%%%%%%%%%%%%%%
\begin{figure}
\includegraphics[width=\columnwidth]{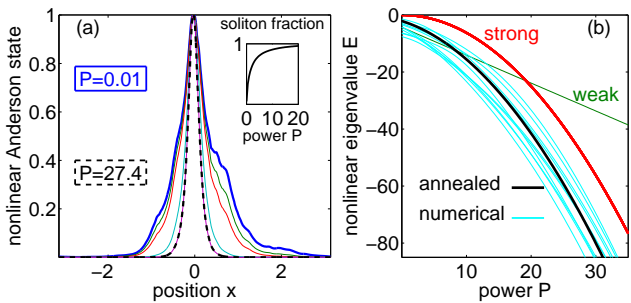}
\caption{(Color online) (a) Plot of the nonlinear Anderson states $|\varphi|/max(|\varphi|)$ for different $P$   ($V_0=4$), two values of the powers are indicated corresponding to the blue and dashed lines;
the inset shows the projection on the soliton profile;
(b) nonlinear eigenvalue $E$ versus $P$ ($V_0=4$) for several disorder realizations (cyan thin lines), compared with the strong (red thick line) and weak (green thick line, only shown for a single realization) expansions, and with the result from the annealed phase-space variational approach (black thick line).
\label{fig_bright}}
\end{figure}
%%%%%%%%%%%%%%%%%%%%
\\\noindent{\it Nonlinear dressing ---}
We detail the transition from the linear Anderson states to the solitary wave.
We limit to the focusing case $\chi=1$ hereafter (the case $\chi=-1$ will be reported elsewhere).
We numerically solve equation (\ref{eq:nlbound}), which is invariant with respect to the scaling $x\rightarrow x/x_0$,
%$z\rightarrow z/x_0^2$,
 $V\rightarrow V x_0^2$ and $\psi\rightarrow\psi/x_0^2$,
such that we can limit the size of system to $x\in [-\pi,\pi]$,
over which periodical boundary conditions are enforced.
The prolongation of the linear states to the nonlinear case is not trivial.
We start from a linear localized state ($\chi=0$) and we prolong to $\chi>0$ by a Newton-Raphson algorithm; 
then, for each $\chi$, we rescale $\fii$, by using the scaling properties of Eq.(\ref{nls1}), so that it corresponds to $\chi=1$,
and we calculate $P$, $H$, $l$ and $E$.
In figure \ref{fig_bright}a we show the shape of the fundamental solution (lowest negative eigenvalue) for increasing power $P$.  The inset shows the projection of the numerically retrieved nonlinear localization with the fundamental soliton {\it sech} profile: as $P$ increases the shape of the disorder induced localization is progressively similar to the soliton.
The trend of the eigenvalue $E$ is shown in Fig.\ref{fig_bright}b, compared with the weak (low $P$) and strong expansions (high $P$), and with Eq.(\ref{EZ}) below: for low $P$ we have a linear trend, Eq.(\ref{eq:weakE}), while the trend follows the solitonic one $E_S=-P^2/16$ for high $P$.
%%%%%%%%%%%%%%%%%%%%
\begin{figure}
\includegraphics[width=\columnwidth]{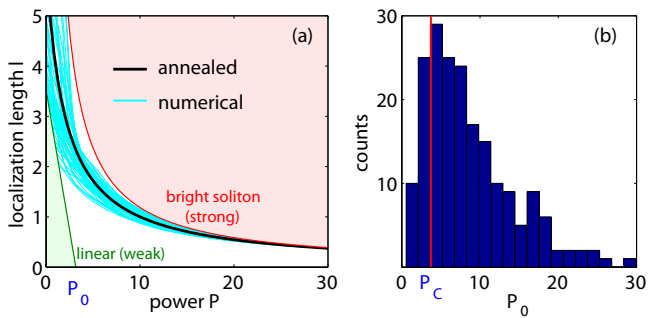}
\caption{(Color online) 
(a) Localization length versus $P$ ($V_0=1$): the results after various disorder realizations are shown (cyan thin lines) and compared with the weak (green line, only shown for a single realization with the corresponding $P_0$ indicated), with the strong expansion (red thick line), and with annealed phase-space approach (black thick line); (b) distribution of the critical power $P_0$ ($200$ disorder realizations), the red vertical line is the analytical result for $P_C$, as given in the text ($V_0=2$).
\label{fig_localization}}
\end{figure}
%%%%%%%%%%%%%%%%%%%%
\\\noindent In Fig.\ref{fig_localization}a we show the calculated localization length compared with the strong perturbation theory and, for a single realization, with the weak perturbation theory, with $P_0$ given by the intercept with the horizontal axis. When 
$P$ increases, the localization length deviates from the linear trend, and follows the bright soliton $l_S=12/P$ at high $P$ for all the considered realizations. 
\\\noindent {\it The phase-space/variational approach ---}
In the weak expansion Eq.(\ref{lweak}), valid as $P\rightarrow 0$, the power $P_0$ for the transition to the soliton depends on the realization of the disorder and has a statistical distribution shown in Fig.\ref{fig_localization}b. The histogram of $P_0$ in Fig.\ref{fig_localization}b is found not to substantially change for a number of realizations greater than $100$.
\\On the other hand, the strong expansion Eq.(\ref{eq:strong}), valid as $P\rightarrow\infty$, completely neglects the
random potential, and the result is independent of the strength of disorder.
Here we introduce an approach based on the statistical mechanics of disordered systems \cite{Conti2011} valid at any order in $P$. 
The first step is to define an appropriate measure based on the fact that the nonlinear Anderson state (\ref{eq:nlbound}) maximizes a weight in the space of all the possible $\psi$. 
Following the fact the nonlinear bound states extremize the Lyapunov functional $F$, and that, for these states, $F$ scales like $1/l$ after Eq.(\ref{eq:Fl}), we consider a Boltzmann like weight: $T_l=\exp(-L/l)$ with $L$ determined in the following.
Note that $\exp(-L/l)$ is the transmission of a slab of disordered material with length $L$ and localization length $l$ \cite{LifshitsBook}.
For a specific disorder realization, we introduce the measure
\begin{equation}
\rho[\psi]=\frac{1}{Z}\exp\left(-\frac{L}{l}\right)
\end{equation}
with $Z$ the ``partition function''
\begin{equation}
Z=\int \exp\left(-\frac{L}{l}\right) d[\psi]\text{,}
\end{equation}
such that $\int \rho[\psi] d[\psi]=1$.
The inverse localization length is calculated as an average over the whole functional space of $\psi$ [after Eq.(\ref{eq:Fl})]:
\begin{equation}
\langle \frac{1}{l}\rangle\equiv\int \frac{1}{l} \rho[\psi] d[\psi]=\frac{1}{Z}\int \frac{1}{l} \exp{\left(-\frac{2L F}{P^2}\right)} d[\psi]\text{.}
\label{eq:Z}
\end{equation}
In (\ref{eq:Z}), the solution of (\ref{eq:nlbound}) is that providing the
highest contribution to the weighted average among all the $\psi$.
Our aim is to find an equation for such a state after averaging over the disorder $V(x)$; this averaging is denoted by an over-line:
\begin{equation}
\overline{\langle\frac{1}{l}\rangle}=- \overline{\partial_L log(Z)}\cong-\partial_L log \overline{Z}\text{.}
\label{annealed}
\end{equation}
In (\ref{annealed}) we used the so-called annealed average $\overline{\log(Z)}\cong\log{\overline{Z}}$,\cite{MPVBook} whose validity is to
be confirmed {\it a posteriori}. We find $\overline Z=\int \exp\left(-2 L F_{eff}/P^2 \right) d[\psi]$, being
\begin{equation}
  F_{eff}\equiv\int \left[|\psi_x|^2-\frac{1}{2}(1+\frac{2 L V_0^2}{P^2})|\psi|^4 -E |\psi|^2\right]dx \text{.}
\end{equation}
This effective Lyapunov functional $F_{eff}$ is extremized by
\begin{equation}
-\psi_{xx}-\left(1+\frac{2 L V_0^2}{P^2}\right)|\psi|^2 \psi=E\psi
\label{eq:avgeq}
\end{equation}
with the constraint $P=\int |\psi|^2 dx$, which gives $E$ as a function $P$.
Eq.(\ref{eq:avgeq}) generalizes the strong perturbation limit Eq.(\ref{eq:strong}), retrieved for $V_0=0$ or $P\rightarrow\infty$, to a finite potential $V_0$ and $P$.
\\Eq.(\ref{eq:avgeq}) shows that the role of the disorder is to alter the nonlinear response, namely to increase the strength of the nonlinear coefficient,
such that solitary waves are obtained at smaller power than in the ordered case. Conversely, the {\it linear} localizations can be seen
as the nonlinear Anderson states in the limit of vanishing power, that is a form of solitons only due to disorder. 
In Eq.(\ref{eq:avgeq}), $P$ explicitly appears;
this is due to the fact that the average over disorder introduces nonlocality  \cite{Snyder97} in the model.
In the defocusing case a result similar to (\ref{eq:avgeq}) is found, with a nonlinear coefficient changing sign at high power,
denoting the absence of localization for large $P$, as we will report in future work.
\\\noindent In the focusing case, by using the fundamental {\it sech} soliton of Eq(\ref{eq:avgeq}), we find the corresponding eigenvalue, denoted as $E_C$:
\begin{equation}
E_C=-\frac{P^2}{16}\left(1+\frac{2 L V_0^2}{P^2}\right)^2
\end{equation}
with the localization length $l_C=3/\sqrt{-E_C}$. $E_C$ is the $P$ and $V_0$ dependent eigenvalue for a generic $L$,
and $l_C$ is the corresponding localization length; note that according to this analysis a measurement of $l_C$ directly provides $E_C$. \\In the next step we determine $L$ by imposing the correct asymptotic value in the linear limit:
as $P\rightarrow 0$,  it must be $E_C\rightarrow\overline E_L\cong - V_0^{4/3}/3$, which furnishes $L=2 V_0^{-4/3} P/\sqrt{3}$;  conversely, in the large $P$ limit one recovers the expected expression $E_C\rightarrow E_S=-P^2/16$. 

\noindent Summarizing, we find the for the nonlinear eigenvalue
\begin{equation}
E_C=-\frac{P^2}{16}\left(1+\frac{P_C}{P}\right)^2\text{,}
\label{EZ}
\end{equation}
with the only parameter $P_C=4 V_0^{2/3}/\sqrt{3}$. Correspondingly, the localization length is 
\begin{equation}
l_C=\frac{12/P}{(1+P_C/P)}\text{,}
\label{lZ}
\end{equation}
which also gives $l_S=12/P$ for large $P$, and the weak limit Eq.(\ref{lweak}) with $l_0=12/P_C$ and $P_0=P_C$.
$P_C$ is the {\it critical} power for the transition from the Anderson localizations to the solitons and is determined by the strength of disorder $V_0$.
In the linear limit $P\rightarrow 0$, Eqs.(\ref{lZ}) and (\ref{EZ}) reproduces the known power-independent link between the localization length and energy $E=E_C=-9/l_0^2$ \cite{LifshitsBook}.
We stress that $l_c$ is the localization length of the state that mostly
contribute to $\overline Z$.
\\\noindent In Figures \ref{fig_bright} and \ref{fig_localization}, we compare this theoretical approach with the numerical simulations at any value of $P$; results for various $V_0$ are used to show that quantitative agreement is found in all of the considered cases.
\\\noindent {\it Stability and superlocalization ---}
We consider the stability of the nonlinear localization:
we calculate the eigenvalues of the linearized problem following the Vakhitov and Kolokolov formulation\cite{VKpaper,Rose1988,KivsharBook}. 
We write $\psi=(\varphi+\delta\psi)\exp(-i E t)$, with $\delta\psi=[u(x)+i v(x)]\exp(\Omega t)$, where $\varphi$ is a solution of Eq.(\ref{eq:nlbound}) and $u$ and $v$ are real valued. Eq.(\ref{eq:nlbound}) is linearized as
\begin{equation}
-\Omega^2 u=L_1 L_0 u\text{,}
\label{VK}
\end{equation}
with the operators $L_0=-\partial_x^2-E+V(x)-\varphi(x)^2$ and $L_1=-\partial_x^2-E+V(x)- 3 \varphi(x)^2$. 
The stable (unstable) eigenvalues correspond to $\Omega^2>0$ ($\Omega^2<0$).
As it happens for the standard solitons, the bound state profile is also an eigenvalue of (\ref{VK}),
with $\Omega=0$ due to the gauge invariance; conversely other neutral modes due to translational invariance are lost due to the symmetry breaking potential $V(x)$.
\begin{figure}
\includegraphics[width=\columnwidth]{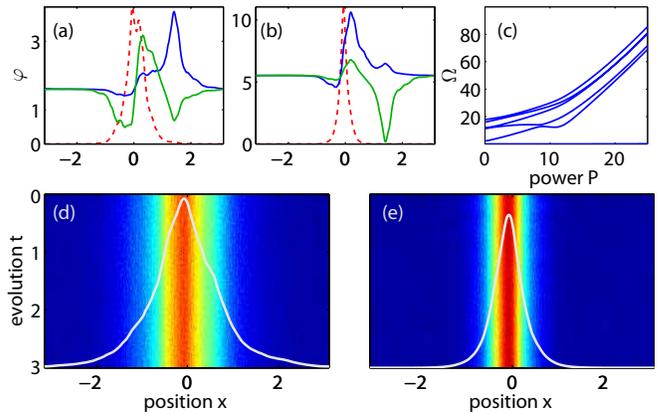}
\caption{(Color online) Stability of the nonlinear Anderson states: (a) dashed, nonlinear state at power $P=7$; full lines, two different superlocalizations arbitrarily vertically shifted  (blu $\Omega=11$,green $\Omega=14$); (b) as in (a) for $P=23$ (blu $\Omega=59$, green $\Omega=61$); (c) eigenvalues versus power $P$ ($V_0=10$);
(d) evolution of the nonlinear states with $P=3.54$ with $10\%$ amplitude noise; (e) as in (d) with $P=13.2$; the white line corresponds to the arbitrarily scaled initial profile $\varphi$ ($V_0=2$).\label{fig_stab}}
\end{figure}
We numerically solve Eq.(\ref{VK}) and find that 
no unstable states are present for the considered disorder realizations and values of $V_0$, demonstrating that the nonlinear Anderson states are indeed
stable with respect to linear perturbations.
The interesting issue is that in regions far from the nonlinear bound states (where $\varphi\cong0$), Eq.(\ref{VK}) still admits non trivial solutions, corresponding to $L_1L_0\cong \mathcal{L}^2=[-\partial_x^2+V(x)-E]^2$,
such that the linear Anderson states correspond to $\Omega^2\cong 0$. 
As $P$ increases, the location of these states drifts towards the center of the nonlinear localization and this coupling results into a power dependent $\Omega^2$ (examples are given in Fig.\ref{fig_stab}a,b,c). These can be taken as ``superlocalizations'' due to interplay between disorder and solitons.
The stability of the nonlinear Anderson states is also verified by their numerically calculated  $t$-evolutions in the presence of a perturbation, as shown in 
Fig.\ref{fig_stab}d,e. 
\\\noindent{\it Conclusions ---} 
We reported on a theoretical approach on nonlinear Anderson localization demonstrating the strong connection between solitons and disorder induced localization.
By a variational formulation we derived closed formulae for the fundamental state providing the trend of the nonlinear eigenvalue and the localization length at any power level in quantitative agreement with numerical simulations. 
Disorder averaged nonlinear Anderson localization 
is found to obey a nonlocal Schroedinger equation 
with a disorder dependent nonlinearity.
Such an equation, in the linear limit, reproduces the linear Anderson states.
The reported approach can be extended to the multidimensional case and to other nonlinearities.
\\ The research leading to these results has received funding from the
European Research Council under the European Community's Seventh Framework Program 
(FP7/2007-2013)/ERC grant agreement n.201766, project {\it Light and Complexity} (COMPLEXLIGHT).
We acknowledge support from the Humboldt foundation.

%\bibliography{MEGAbib}

\begin{thebibliography}{41}
\expandafter\ifx\csname natexlab\endcsname\relax\def\natexlab#1{#1}\fi
\expandafter\ifx\csname bibnamefont\endcsname\relax
  \def\bibnamefont#1{#1}\fi
\expandafter\ifx\csname bibfnamefont\endcsname\relax
  \def\bibfnamefont#1{#1}\fi
\expandafter\ifx\csname citenamefont\endcsname\relax
  \def\citenamefont#1{#1}\fi
\expandafter\ifx\csname url\endcsname\relax
  \def\url#1{\texttt{#1}}\fi
\expandafter\ifx\csname urlprefix\endcsname\relax\def\urlprefix{URL }\fi
\providecommand{\bibinfo}[2]{#2}
\providecommand{\eprint}[2][]{\url{#2}}

\bibitem[{\citenamefont{Zabusky and Kruskal}(1965)}]{ZK65}
\bibinfo{author}{\bibfnamefont{N.~J.} \bibnamefont{Zabusky}} \bibnamefont{and}
  \bibinfo{author}{\bibfnamefont{M.~D.} \bibnamefont{Kruskal}},
  \bibinfo{journal}{Phys. Rev. Lett.} \textbf{\bibinfo{volume}{15}},
  \bibinfo{pages}{240} (\bibinfo{year}{1965}).

\bibitem[{\citenamefont{Anderson}(1958)}]{Anderson58}
\bibinfo{author}{\bibfnamefont{P.}~\bibnamefont{Anderson}},
  \bibinfo{journal}{Phys. Rev.} \textbf{\bibinfo{volume}{109}},
  \bibinfo{pages}{1492} (\bibinfo{year}{1958}).

\bibitem[{\citenamefont{Drazin and Johnson}(1989)}]{DrazinBook}
\bibinfo{author}{\bibfnamefont{P.~G.} \bibnamefont{Drazin}} \bibnamefont{and}
  \bibinfo{author}{\bibfnamefont{R.~S.} \bibnamefont{Johnson}},
  \emph{\bibinfo{title}{Solitons: An Introduction}}
  (\bibinfo{publisher}{Cambridge University Press}, \bibinfo{address}{New
  York}, \bibinfo{year}{1989}).

\bibitem[{\citenamefont{Lifshits et~al.}(1988)\citenamefont{Lifshits,
  Gredeskul, and Pastur}}]{LifshitsBook}
\bibinfo{author}{\bibfnamefont{I.~M.} \bibnamefont{Lifshits}},
  \bibinfo{author}{\bibfnamefont{S.~A.} \bibnamefont{Gredeskul}},
  \bibnamefont{and} \bibinfo{author}{\bibfnamefont{L.~A.}
  \bibnamefont{Pastur}}, \emph{\bibinfo{title}{Introduction to the theory of
  disorder systems}} (\bibinfo{publisher}{John Wiley \& Sons},
  \bibinfo{year}{1988}).

\bibitem[{\citenamefont{Kivshar et~al.}(1990)\citenamefont{Kivshar, Gredeskul,
  S\'anchez, and V\'azquez}}]{Kivshar1990}
\bibinfo{author}{\bibfnamefont{Y.~S.} \bibnamefont{Kivshar}},
  \bibinfo{author}{\bibfnamefont{S.~A.} \bibnamefont{Gredeskul}},
  \bibinfo{author}{\bibfnamefont{A.}~\bibnamefont{S\'anchez}},
  \bibnamefont{and}
  \bibinfo{author}{\bibfnamefont{L.}~\bibnamefont{V\'azquez}},
  \bibinfo{journal}{Phys. Rev. Lett.} \textbf{\bibinfo{volume}{64}},
  \bibinfo{pages}{1693} (\bibinfo{year}{1990}).

\bibitem[{\citenamefont{Zavt et~al.}(1993)\citenamefont{Zavt, Wagner, and
  L{\"u}tze}}]{Zavt93}
\bibinfo{author}{\bibfnamefont{G.~S.} \bibnamefont{Zavt}},
  \bibinfo{author}{\bibfnamefont{M.}~\bibnamefont{Wagner}}, \bibnamefont{and}
  \bibinfo{author}{\bibfnamefont{A.}~\bibnamefont{L{\"u}tze}},
  \bibinfo{journal}{Phys. Rev. E} \textbf{\bibinfo{volume}{47}},
  \bibinfo{pages}{4108} (\bibinfo{year}{1993}).

\bibitem[{\citenamefont{Sukhorukov et~al.}(2001)\citenamefont{Sukhorukov,
  Kivshar, Bang, Rasmussen, and Christiansen}}]{Sukh2001}
\bibinfo{author}{\bibfnamefont{A.~A.} \bibnamefont{Sukhorukov}},
  \bibinfo{author}{\bibfnamefont{Y.~S.} \bibnamefont{Kivshar}},
  \bibinfo{author}{\bibfnamefont{O.}~\bibnamefont{Bang}},
  \bibinfo{author}{\bibfnamefont{J.~J.} \bibnamefont{Rasmussen}},
  \bibnamefont{and} \bibinfo{author}{\bibfnamefont{P.~L.}
  \bibnamefont{Christiansen}}, \bibinfo{journal}{Phys. Rev. E}
  \textbf{\bibinfo{volume}{63}}, \bibinfo{pages}{036601}
  (\bibinfo{year}{2001}).

\bibitem[{\citenamefont{Staliunas}(2003)}]{Staliunas03}
\bibinfo{author}{\bibfnamefont{K.}~\bibnamefont{Staliunas}},
  \bibinfo{journal}{Phys. Rev. A} \textbf{\bibinfo{volume}{68}},
  \bibinfo{pages}{013801} (\bibinfo{year}{2003}).

\bibitem[{\citenamefont{Bliokh et~al.}(2006)\citenamefont{Bliokh, Bliokh,
  Freilikher, Genack, Hu, and Sebbah}}]{bliokh06}
\bibinfo{author}{\bibfnamefont{K.~Y.} \bibnamefont{Bliokh}},
  \bibinfo{author}{\bibfnamefont{Y.~P.} \bibnamefont{Bliokh}},
  \bibinfo{author}{\bibfnamefont{V.}~\bibnamefont{Freilikher}},
  \bibinfo{author}{\bibfnamefont{A.~Z.} \bibnamefont{Genack}},
  \bibinfo{author}{\bibfnamefont{B.}~\bibnamefont{Hu}}, \bibnamefont{and}
  \bibinfo{author}{\bibfnamefont{P.}~\bibnamefont{Sebbah}},
  \bibinfo{journal}{Phys. Rev. Lett.} \textbf{\bibinfo{volume}{97}},
  \bibinfo{pages}{243904} (\bibinfo{year}{2006}).

\bibitem[{\citenamefont{Bliokh et~al.}(2008)\citenamefont{Bliokh, Bliokh,
  Freilikher, Savel'ev, and Nori}}]{bliokh08}
\bibinfo{author}{\bibfnamefont{K.~Y.} \bibnamefont{Bliokh}},
  \bibinfo{author}{\bibfnamefont{Y.~P.} \bibnamefont{Bliokh}},
  \bibinfo{author}{\bibfnamefont{V.}~\bibnamefont{Freilikher}},
  \bibinfo{author}{\bibfnamefont{S.}~\bibnamefont{Savel'ev}}, \bibnamefont{and}
  \bibinfo{author}{\bibfnamefont{F.}~\bibnamefont{Nori}},
  \bibinfo{journal}{Rev. Mod. Phys.} \textbf{\bibinfo{volume}{80}},
  \bibinfo{pages}{1201} (\bibinfo{year}{2008}).

\bibitem[{\citenamefont{Shadrivov et~al.}(2010)\citenamefont{Shadrivov, Bliokh,
  Bliokh, Freilikher, and Kivshar}}]{Kivshar10}
\bibinfo{author}{\bibfnamefont{I.~V.} \bibnamefont{Shadrivov}},
  \bibinfo{author}{\bibfnamefont{K.~Y.} \bibnamefont{Bliokh}},
  \bibinfo{author}{\bibfnamefont{Y.~P.} \bibnamefont{Bliokh}},
  \bibinfo{author}{\bibfnamefont{V.}~\bibnamefont{Freilikher}},
  \bibnamefont{and} \bibinfo{author}{\bibfnamefont{Y.~S.}
  \bibnamefont{Kivshar}}, \bibinfo{journal}{Phys. Rev. Lett.}
  \textbf{\bibinfo{volume}{104}}, \bibinfo{pages}{123902}
  (\bibinfo{year}{2010}).

\bibitem[{\citenamefont{Folli and Conti}(2010)}]{Folli10}
\bibinfo{author}{\bibfnamefont{V.}~\bibnamefont{Folli}} \bibnamefont{and}
  \bibinfo{author}{\bibfnamefont{C.}~\bibnamefont{Conti}},
  \bibinfo{journal}{Phys. Rev. Lett.} \textbf{\bibinfo{volume}{104}},
  \bibinfo{pages}{193901} (\bibinfo{year}{2010}).

\bibitem[{\citenamefont{Jovic et~al.}(2011)\citenamefont{Jovic, Belic, and
  Denz}}]{Denz2011}
\bibinfo{author}{\bibfnamefont{D.~M.} \bibnamefont{Jovic}},
  \bibinfo{author}{\bibfnamefont{M.~R.} \bibnamefont{Belic}}, \bibnamefont{and}
  \bibinfo{author}{\bibfnamefont{C.}~\bibnamefont{Denz}},
  \bibinfo{journal}{Phys. Rev. A} \textbf{\bibinfo{volume}{84}},
  \bibinfo{pages}{043811} (\bibinfo{year}{2011}).

\bibitem[{\citenamefont{Folli and Conti}(2011)}]{Folli11}
\bibinfo{author}{\bibfnamefont{V.}~\bibnamefont{Folli}} \bibnamefont{and}
  \bibinfo{author}{\bibfnamefont{C.}~\bibnamefont{Conti}},
  \bibinfo{journal}{Opt. Lett.} \textbf{\bibinfo{volume}{36}},
  \bibinfo{pages}{2830} (\bibinfo{year}{2011}).

\bibitem[{\citenamefont{Folli and Conti}(2012)}]{Folli12}
\bibinfo{author}{\bibfnamefont{V.}~\bibnamefont{Folli}} \bibnamefont{and}
  \bibinfo{author}{\bibfnamefont{C.}~\bibnamefont{Conti}},
  \bibinfo{journal}{Opt. Lett.} \textbf{\bibinfo{volume}{37}},
  \bibinfo{pages}{332} (\bibinfo{year}{2012}).

\bibitem[{\citenamefont{{Maucher} et~al.}(2012)\citenamefont{{Maucher},
  {Krolikowski}, and {Skupin}}}]{Skupin2012}
\bibinfo{author}{\bibfnamefont{F.}~\bibnamefont{{Maucher}}},
  \bibinfo{author}{\bibfnamefont{W.}~\bibnamefont{{Krolikowski}}},
  \bibnamefont{and} \bibinfo{author}{\bibfnamefont{S.}~\bibnamefont{{Skupin}}},
  \bibinfo{journal}{ArXiv e-prints}  (\bibinfo{year}{2012}),
  \eprint{1202.2074}.

\bibitem[{\citenamefont{Sacha et~al.}(2009)\citenamefont{Sacha, M{\"u}ller,
  Delande, and Zakrzewski}}]{Sacha09}
\bibinfo{author}{\bibfnamefont{K.}~\bibnamefont{Sacha}},
  \bibinfo{author}{\bibfnamefont{C.~A.} \bibnamefont{M{\"u}ller}},
  \bibinfo{author}{\bibfnamefont{D.}~\bibnamefont{Delande}}, \bibnamefont{and}
  \bibinfo{author}{\bibfnamefont{J.}~\bibnamefont{Zakrzewski}},
  \bibinfo{journal}{Phys. Rev. Lett.} \textbf{\bibinfo{volume}{103}},
  \bibinfo{eid}{210402} (\bibinfo{year}{2009}).

\bibitem[{\citenamefont{Kartashov et~al.}(2008)\citenamefont{Kartashov,
  Vysloukh, and Torner}}]{kartashov08}
\bibinfo{author}{\bibfnamefont{Y.~V.} \bibnamefont{Kartashov}},
  \bibinfo{author}{\bibfnamefont{V.~A.} \bibnamefont{Vysloukh}},
  \bibnamefont{and} \bibinfo{author}{\bibfnamefont{L.}~\bibnamefont{Torner}},
  \bibinfo{journal}{Phys. Rev. A} \textbf{\bibinfo{volume}{77}},
  \bibinfo{pages}{051802} (\bibinfo{year}{2008}).

\bibitem[{\citenamefont{Modugno}(2010)}]{Modugno2010}
\bibinfo{author}{\bibfnamefont{G.}~\bibnamefont{Modugno}},
  \bibinfo{journal}{Repors on Progress in Physics}
  \textbf{\bibinfo{volume}{73}}, \bibinfo{pages}{102401}
  (\bibinfo{year}{2010}).

\bibitem[{\citenamefont{Schwartz et~al.}(2007)\citenamefont{Schwartz, Bartal,
  Fishman, and Segev}}]{Swartz07}
\bibinfo{author}{\bibfnamefont{T.}~\bibnamefont{Schwartz}},
  \bibinfo{author}{\bibfnamefont{G.}~\bibnamefont{Bartal}},
  \bibinfo{author}{\bibfnamefont{S.}~\bibnamefont{Fishman}}, \bibnamefont{and}
  \bibinfo{author}{\bibfnamefont{M.}~\bibnamefont{Segev}},
  \bibinfo{journal}{Nature} \textbf{\bibinfo{volume}{446}}, \bibinfo{pages}{52}
  (\bibinfo{year}{2007}).

\bibitem[{\citenamefont{Conti and Fratalocchi}(2008)}]{Conti08PhC}
\bibinfo{author}{\bibfnamefont{C.}~\bibnamefont{Conti}} \bibnamefont{and}
  \bibinfo{author}{\bibfnamefont{A.}~\bibnamefont{Fratalocchi}},
  \bibinfo{journal}{Nat. Physics} \textbf{\bibinfo{volume}{4}},
  \bibinfo{pages}{794} (\bibinfo{year}{2008}).

\bibitem[{\citenamefont{{El-Dardiry} et~al.}(2012)\citenamefont{{El-Dardiry},
  {Faez}, and {Lagendijk}}}]{Lagendijk2012}
\bibinfo{author}{\bibfnamefont{R.~G.~S.} \bibnamefont{{El-Dardiry}}},
  \bibinfo{author}{\bibfnamefont{S.}~\bibnamefont{{Faez}}}, \bibnamefont{and}
  \bibinfo{author}{\bibfnamefont{A.}~\bibnamefont{{Lagendijk}}},
  \bibinfo{journal}{ArXiv e-prints}  (\bibinfo{year}{2012}),
  \eprint{1201.0635}.

\bibitem[{\citenamefont{Bodyfelt et~al.}(2010)\citenamefont{Bodyfelt, Kottos,
  and Shapiro}}]{Shapiro2010}
\bibinfo{author}{\bibfnamefont{J.~D.} \bibnamefont{Bodyfelt}},
  \bibinfo{author}{\bibfnamefont{T.}~\bibnamefont{Kottos}}, \bibnamefont{and}
  \bibinfo{author}{\bibfnamefont{B.}~\bibnamefont{Shapiro}},
  \bibinfo{journal}{Phys. Rev. Lett.} \textbf{\bibinfo{volume}{104}},
  \bibinfo{pages}{164102} (\bibinfo{year}{2010}).

\bibitem[{\citenamefont{Paul et~al.}(2007)\citenamefont{Paul, Schlagheck,
  Leboeuf, and Pavloff}}]{Paul2007}
\bibinfo{author}{\bibfnamefont{T.}~\bibnamefont{Paul}},
  \bibinfo{author}{\bibfnamefont{P.}~\bibnamefont{Schlagheck}},
  \bibinfo{author}{\bibfnamefont{P.}~\bibnamefont{Leboeuf}}, \bibnamefont{and}
  \bibinfo{author}{\bibfnamefont{N.}~\bibnamefont{Pavloff}},
  \bibinfo{journal}{Phys. Rev. Lett.} \textbf{\bibinfo{volume}{98}},
  \bibinfo{pages}{210602} (\bibinfo{year}{2007}).

\bibitem[{\citenamefont{Billy et~al.}(2008)\citenamefont{Billy, Josse, Zuo,
  Bernard, Hambrecht, Lugan, Clement, Sanchez-Palencia, Bouyer, and
  Aspect}}]{AspectNature2008}
\bibinfo{author}{\bibfnamefont{J.}~\bibnamefont{Billy}},
  \bibinfo{author}{\bibfnamefont{V.}~\bibnamefont{Josse}},
  \bibinfo{author}{\bibfnamefont{Z.}~\bibnamefont{Zuo}},
  \bibinfo{author}{\bibfnamefont{A.}~\bibnamefont{Bernard}},
  \bibinfo{author}{\bibfnamefont{B.}~\bibnamefont{Hambrecht}},
  \bibinfo{author}{\bibfnamefont{P.}~\bibnamefont{Lugan}},
  \bibinfo{author}{\bibfnamefont{D.}~\bibnamefont{Clement}},
  \bibinfo{author}{\bibfnamefont{L.}~\bibnamefont{Sanchez-Palencia}},
  \bibinfo{author}{\bibfnamefont{P.}~\bibnamefont{Bouyer}}, \bibnamefont{and}
  \bibinfo{author}{\bibfnamefont{A.}~\bibnamefont{Aspect}},
  \bibinfo{journal}{Nature} \textbf{\bibinfo{volume}{453}},
  \bibinfo{pages}{891} (\bibinfo{year}{2008}).

\bibitem[{\citenamefont{Roati et~al.}(2008)\citenamefont{Roati, D'Errico,
  Fallani, Fattori, Fort, Zaccanti, Modugno, Modugno, and
  Inguscio}}]{InguscioNature2008}
\bibinfo{author}{\bibfnamefont{G.}~\bibnamefont{Roati}},
  \bibinfo{author}{\bibfnamefont{C.}~\bibnamefont{D'Errico}},
  \bibinfo{author}{\bibfnamefont{L.}~\bibnamefont{Fallani}},
  \bibinfo{author}{\bibfnamefont{M.}~\bibnamefont{Fattori}},
  \bibinfo{author}{\bibfnamefont{C.}~\bibnamefont{Fort}},
  \bibinfo{author}{\bibfnamefont{M.}~\bibnamefont{Zaccanti}},
  \bibinfo{author}{\bibfnamefont{G.}~\bibnamefont{Modugno}},
  \bibinfo{author}{\bibfnamefont{M.}~\bibnamefont{Modugno}}, \bibnamefont{and}
  \bibinfo{author}{\bibfnamefont{M.}~\bibnamefont{Inguscio}},
  \bibinfo{journal}{Nature} \textbf{\bibinfo{volume}{453}},
  \bibinfo{pages}{895} (\bibinfo{year}{2008}).

\bibitem[{\citenamefont{Leonetti et~al.}(2012)\citenamefont{Leonetti, Conti,
  and Lopez}}]{Leonetti2012}
\bibinfo{author}{\bibfnamefont{M.}~\bibnamefont{Leonetti}},
  \bibinfo{author}{\bibfnamefont{C.}~\bibnamefont{Conti}}, \bibnamefont{and}
  \bibinfo{author}{\bibfnamefont{C.}~\bibnamefont{Lopez}},
  \bibinfo{journal}{Applied Physics Letters} \textbf{\bibinfo{volume}{101}},
  \bibinfo{pages}{051104} (\bibinfo{year}{2012}).

\bibitem[{\citenamefont{Hernandez-Figueroa
  et~al.}(2008)\citenamefont{Hernandez-Figueroa, Zamboni-Rached, and
  E.}}]{RecamiBook}
\bibinfo{editor}{\bibfnamefont{H.~E.} \bibnamefont{Hernandez-Figueroa}},
  \bibinfo{editor}{\bibfnamefont{M.}~\bibnamefont{Zamboni-Rached}},
  \bibnamefont{and} \bibinfo{editor}{\bibfnamefont{R.}~\bibnamefont{E.}}, eds.,
  \emph{\bibinfo{title}{Localized Waves}} (\bibinfo{publisher}{John Wiley \&
  Sons}, \bibinfo{year}{2008}).

\bibitem[{\citenamefont{Conti}(2004)}]{Conti04c}
\bibinfo{author}{\bibfnamefont{C.}~\bibnamefont{Conti}},
  \bibinfo{journal}{Phys. Rev. E} \textbf{\bibinfo{volume}{70}},
  \bibinfo{eid}{046613} (pages~\bibinfo{numpages}{12}) (\bibinfo{year}{2004}).

\bibitem[{\citenamefont{Kivshar and Agrawal}(2003)}]{KivsharBook}
\bibinfo{author}{\bibfnamefont{Y.}~\bibnamefont{Kivshar}} \bibnamefont{and}
  \bibinfo{author}{\bibfnamefont{G.~P.} \bibnamefont{Agrawal}},
  \emph{\bibinfo{title}{Optical solitons}} (\bibinfo{publisher}{Academic
  Press}, \bibinfo{address}{New York}, \bibinfo{year}{2003}).

\bibitem[{\citenamefont{Gredeskul and Kivshar}(1992)}]{Gred92}
\bibinfo{author}{\bibfnamefont{S.~A.} \bibnamefont{Gredeskul}}
  \bibnamefont{and} \bibinfo{author}{\bibfnamefont{Y.}~\bibnamefont{Kivshar}},
  \bibinfo{journal}{Physics Reports} \textbf{\bibinfo{volume}{1}},
  \bibinfo{pages}{1} (\bibinfo{year}{1992}).

\bibitem[{\citenamefont{Kopidakis and Aubry}(2000)}]{Kopidakis2000}
\bibinfo{author}{\bibfnamefont{G.}~\bibnamefont{Kopidakis}} \bibnamefont{and}
  \bibinfo{author}{\bibfnamefont{S.}~\bibnamefont{Aubry}},
  \bibinfo{journal}{Phys. Rev. Lett.} \textbf{\bibinfo{volume}{84}},
  \bibinfo{pages}{3236} (\bibinfo{year}{2000}).

\bibitem[{\citenamefont{Pikovsky and Shepelyansky}(2008)}]{Pikovsky2008}
\bibinfo{author}{\bibfnamefont{A.~S.} \bibnamefont{Pikovsky}} \bibnamefont{and}
  \bibinfo{author}{\bibfnamefont{D.~L.} \bibnamefont{Shepelyansky}},
  \bibinfo{journal}{Phys. Rev. Lett.} \textbf{\bibinfo{volume}{100}},
  \bibinfo{pages}{094101} (\bibinfo{year}{2008}).

\bibitem[{\citenamefont{Flach et~al.}(2009)\citenamefont{Flach, Krimer, and
  Skokos}}]{Flach2009}
\bibinfo{author}{\bibfnamefont{S.}~\bibnamefont{Flach}},
  \bibinfo{author}{\bibfnamefont{D.~O.} \bibnamefont{Krimer}},
  \bibnamefont{and} \bibinfo{author}{\bibfnamefont{C.}~\bibnamefont{Skokos}},
  \bibinfo{journal}{Phys. Rev. Lett.} \textbf{\bibinfo{volume}{102}},
  \bibinfo{pages}{024101} (\bibinfo{year}{2009}).

\bibitem[{\citenamefont{Fishman et~al.}(2012)\citenamefont{Fishman, Krivolapov,
  and Soffer}}]{Fishman2012}
\bibinfo{author}{\bibfnamefont{S.}~\bibnamefont{Fishman}},
  \bibinfo{author}{\bibfnamefont{Y.}~\bibnamefont{Krivolapov}},
  \bibnamefont{and} \bibinfo{author}{\bibfnamefont{A.}~\bibnamefont{Soffer}},
  \bibinfo{journal}{Nonlinearity} \textbf{\bibinfo{volume}{25}},
  \bibinfo{pages}{R53} (\bibinfo{year}{2012}).

\bibitem[{\citenamefont{Snyder and Mitchell}(1997)}]{Snyder97}
\bibinfo{author}{\bibfnamefont{A.~W.} \bibnamefont{Snyder}} \bibnamefont{and}
  \bibinfo{author}{\bibfnamefont{D.~J.} \bibnamefont{Mitchell}},
  \bibinfo{journal}{Science} \textbf{\bibinfo{volume}{276}},
  \bibinfo{pages}{1538} (\bibinfo{year}{1997}).

\bibitem[{\citenamefont{Halperin}(1965)}]{Halperin65}
\bibinfo{author}{\bibfnamefont{B.~I.} \bibnamefont{Halperin}},
  \bibinfo{journal}{Phys. Rev.} \textbf{\bibinfo{volume}{139}},
  \bibinfo{pages}{A104} (\bibinfo{year}{1965}).

\bibitem[{\citenamefont{Conti and Leuzzi}(2011)}]{Conti2011}
\bibinfo{author}{\bibfnamefont{C.}~\bibnamefont{Conti}} \bibnamefont{and}
  \bibinfo{author}{\bibfnamefont{L.}~\bibnamefont{Leuzzi}},
  \bibinfo{journal}{Phys. Rev. B} \textbf{\bibinfo{volume}{83}},
  \bibinfo{pages}{134204} (\bibinfo{year}{2011}).

\bibitem[{\citenamefont{M{\'e}zard et~al.}(1987)\citenamefont{M{\'e}zard,
  Parisi, and Virasoro}}]{MPVBook}
\bibinfo{author}{\bibfnamefont{M.}~\bibnamefont{M{\'e}zard}},
  \bibinfo{author}{\bibfnamefont{G.}~\bibnamefont{Parisi}}, \bibnamefont{and}
  \bibinfo{author}{\bibfnamefont{M.~A.} \bibnamefont{Virasoro}},
  \emph{\bibinfo{title}{Spin glass theory and beyond}}
  (\bibinfo{publisher}{World Scientific}, \bibinfo{address}{Singapore},
  \bibinfo{year}{1987}).

\bibitem[{\citenamefont{Vakhitov and Kolokolov}(1973)}]{VKpaper}
\bibinfo{author}{\bibfnamefont{N.~G.} \bibnamefont{Vakhitov}} \bibnamefont{and}
  \bibinfo{author}{\bibfnamefont{A.~A.} \bibnamefont{Kolokolov}},
  \bibinfo{journal}{Radiophys. Quantum Electron.}
  \textbf{\bibinfo{volume}{16}}, \bibinfo{pages}{783} (\bibinfo{year}{1973}).

\bibitem[{\citenamefont{Rose and Weinstein}(1988)}]{Rose1988}
\bibinfo{author}{\bibfnamefont{H.~A.} \bibnamefont{Rose}} \bibnamefont{and}
  \bibinfo{author}{\bibfnamefont{M.~I.} \bibnamefont{Weinstein}},
  \bibinfo{journal}{Physica D} \textbf{\bibinfo{volume}{30}},
  \bibinfo{pages}{207} (\bibinfo{year}{1988}).

\end{thebibliography}

\end{document}